\documentstyle[11pt,newpasp,twoside]{article}
\def\edcomment#1{\iffalse\marginpar{\raggedright\sl#1\/}\else\relax\fi}
\reversemarginpar

\begin{document}
\title{The Rest of the Story: Radio Pulsars and IR through 
$\gamma$-ray Emission}
 \author{Roger W. Romani}
\affil{Dept. of Physics, Stanford University, Stanford, CA 94305-4060}

\begin{abstract}
Recent observations have detected a number of young pulsars
from the power peak in the $\gamma$-ray band to the incoherent photon
peak in the optical/IR.  We have made progress on the multiwavelength 
phenomenology of pulsar emission and beaming, but a wide variation 
of light curves between different objects and different energy bands
makes the full story complex. I sketch here a
`Unified Model' of pulsar beaming and summarize the radiation
mechanisms and their interplay in outer magnetosphere models of pulsar
high energy emission.
\end{abstract}

\section{Introduction}

        There has been much recent progress in detecting and modeling
high energy ($\sim 10^0 - 10^7$eV) magnetospheric emission from spin powered
pulsars. I will not discuss the coherent radio pulsations
which provide the primary channel for pulsar discovery and 
dynamics/evolutionary studies (as our three honorees and their colleagues 
continue to teach us). The mechanism producing the radio has been very 
difficult to reconstruct (however see Gupta, Jenet, Melrose, these proceedings),
but a well established phenomenology suggests that this is relatively low 
altitude emission above the magnetic polar cap. I will also only briefly describe
the soft X-ray thermal emission from neutron star surfaces (for a recent
review, see Pavlov, Zavlin \& Sanwal 2002). Thus we focus here on the incoherent,
non-thermal radiation, which is the product of magnetosphere gaps.

This radiation is highly pulsed and, hence, highly beamed. With pulse profiles 
having
characteristic component widths $< 0.1$ in phase and the cusp of the Crab
optical pulse extending over no more than $\delta \phi \sim 
0.01$ we must have radiation
dominated by $e^\pm$ with $\Gamma \ge 10-100$. We also must have relatively
narrow radiating zones in the magnetosphere, where again the component widths
restrict us to a few percent of the open field line volume.

\section{Light Curves}

	To set the stage I plot a compilation of light curves from many sources,
updated from Romani (2000). I have placed the closest approach of the radio pole
at phase 0 and 1.5 pulse periods are shown for clarity. 
For Geminga the phase is set by the predicted outer magnetosphere pulses 
matched to the $\gamma$-ray. For PSR B1055-52, the radio polarization indicates 
that the pulse at $\phi=0$ passes closest to the line of sight; the interpulse 
is apparently stronger. In several cases 
the source has only an unpulsed detection; these are shown as flat lines.

\begin{figure}[b!]
\vspace*{8.1cm}
\includegraphics{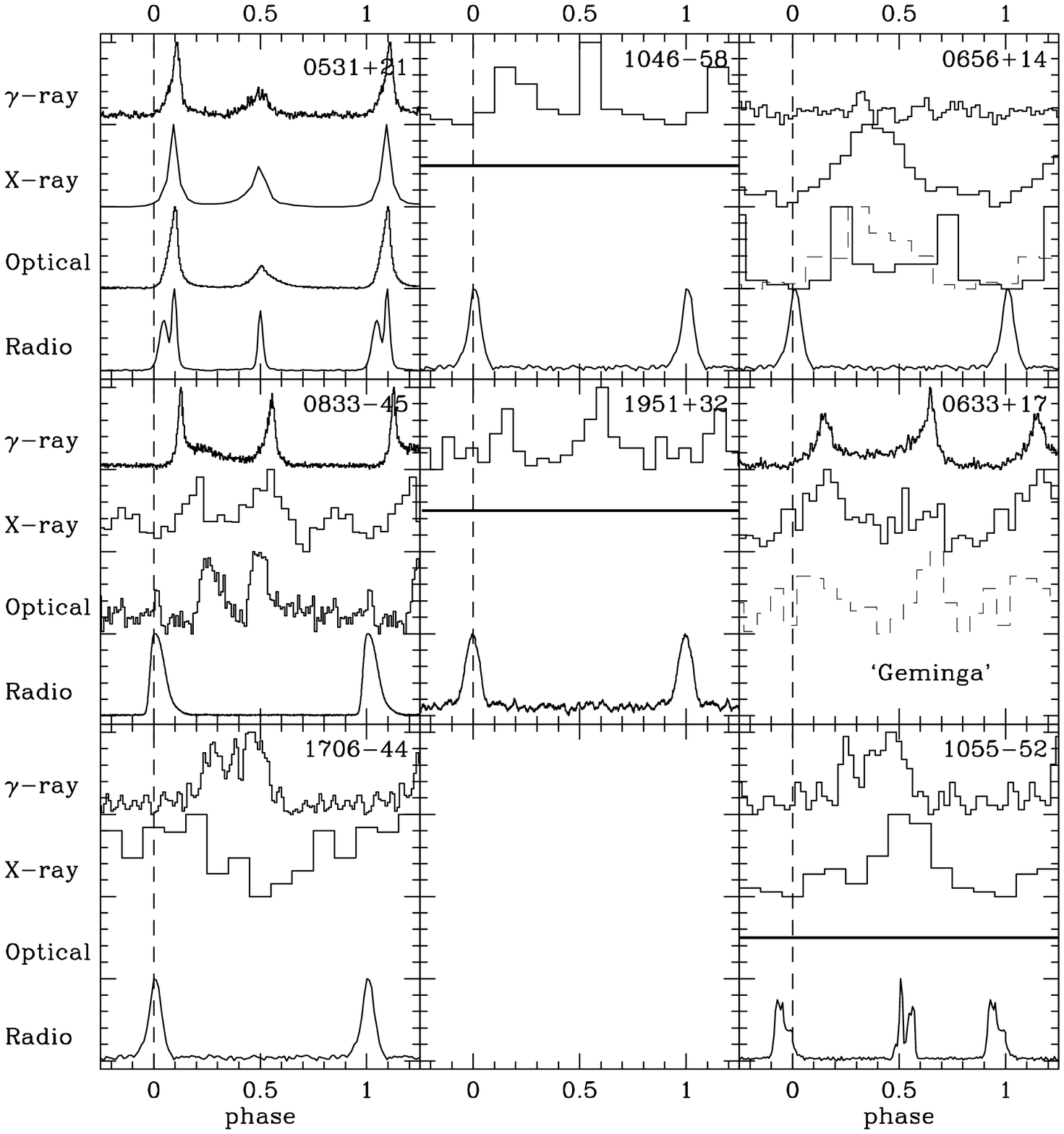}
\caption{Young pulsar light curves (from many literature sources).}
\end{figure}

       In the outer magnetosphere picture
(Morini 1983; Cheng, Ho \& Ruderman 1986) the light curves displayed in this
way make a fairly coherent set. The GeV pulses have double peaks
which come from the pole at $\phi=0.5$ and are beamed toward the spin equator
($\zeta = \pi/2$). Aberration shifts the peaks forward in phase and pulsars
viewed at smaller $\zeta$ have a smaller GeV peak separation; the model
provides a quantitative match to the observed profiles 
(Romani \& Yadigaroglu 1995). The non-thermal
optical and hard X-ray emissions lie largely within the $\gamma$-ray peaks and
arise at lower altitudes, close to the null charge surface that serves as
the inner boundary of the acceleration gap.  The interpretation of the soft
X-ray emission is complicated by the presence of thermal surface flux,
which itself may be pulsed if the surface temperature is non-uniform.
The radio emission, being more tightly beamed,
shows narrow pulses covering a small portion of the sky. The exception
to this picture is the Crab, whose strong non-thermal accelerators seem to
keep pair production rates and particle densities so high
that all emissions (radio-$\gamma$-ray) are dominated by the
outer gap zone.  This interpretation has encouraging successes, but is not unique.
In particular, Harding and colleagues ({\it e.g.} Daugherty and Harding 1996)
have described an aligned rotator polar cap model that can produce
high energy hollow cone emission. 

\section{Physical Processes}

	Light curves locate the position of the radiation zones,
but spectral energy distributions (SEDs) are the key to identifying the
radiation processes.  Thompson (1998) has compiled a useful set of broad-band
SEDs for the young pulsars. The most striking feature of these
$\nu F_\nu$ plots is the power peak at $> 100$MeV,
with a rollover at several GeV. This component
clearly arises from the radiation-reaction limit process for the primary
magnetosphere $e^\pm$. Most recent models ({\it e.g.} Daugherty \& Harding
1996,  Romani 1996, Zhang \& Cheng 1997, Hirotani \& Shibata 1999) conclude
that the particles are curvature radiation limited. 

	At present, outer magnetosphere spectral computations are 
more complex and less complete than near-surface polar cap 
models. The key difference is that at the polar cap 
$\gamma B \longrightarrow e^\pm$ 
provides an efficient source of pair plasma to limit the gap potentials 
and provide the radiating plasma. Moreover, in modern polar cap models
(Harding, Muslimov \& Zhang 2002) the acceleration field is set by general
relativistic frame dragging, providing some insulation from the
details of gap electrodynamics. In contrast, outer gap models produce
pairs via $\gamma-\gamma$, and rely on the charge gaps themselves to provide
the acceleration field. In fact in some pulsars 
the soft target photons are provided by the gap as well. In this way gap
luminosity and spectrum computations are (at least) doubly recursive
(Figure 2).
In effect, outer magnetosphere gaps are self-limiting: to provide the
relatively large GeV efficiencies observed for some pulsars from curvature 
radiation of $\Gamma \sim 10^{7.5}$ $e^\pm$ primaries the gaps must
{\it by definition} be barely closed. Thus computations are difficult and
to get the details right will likely require dynamic 3-D gap models,
well beyond the state of the art today.

\begin{figure}[b!]
\vspace*{4.3cm}
\includegraphics{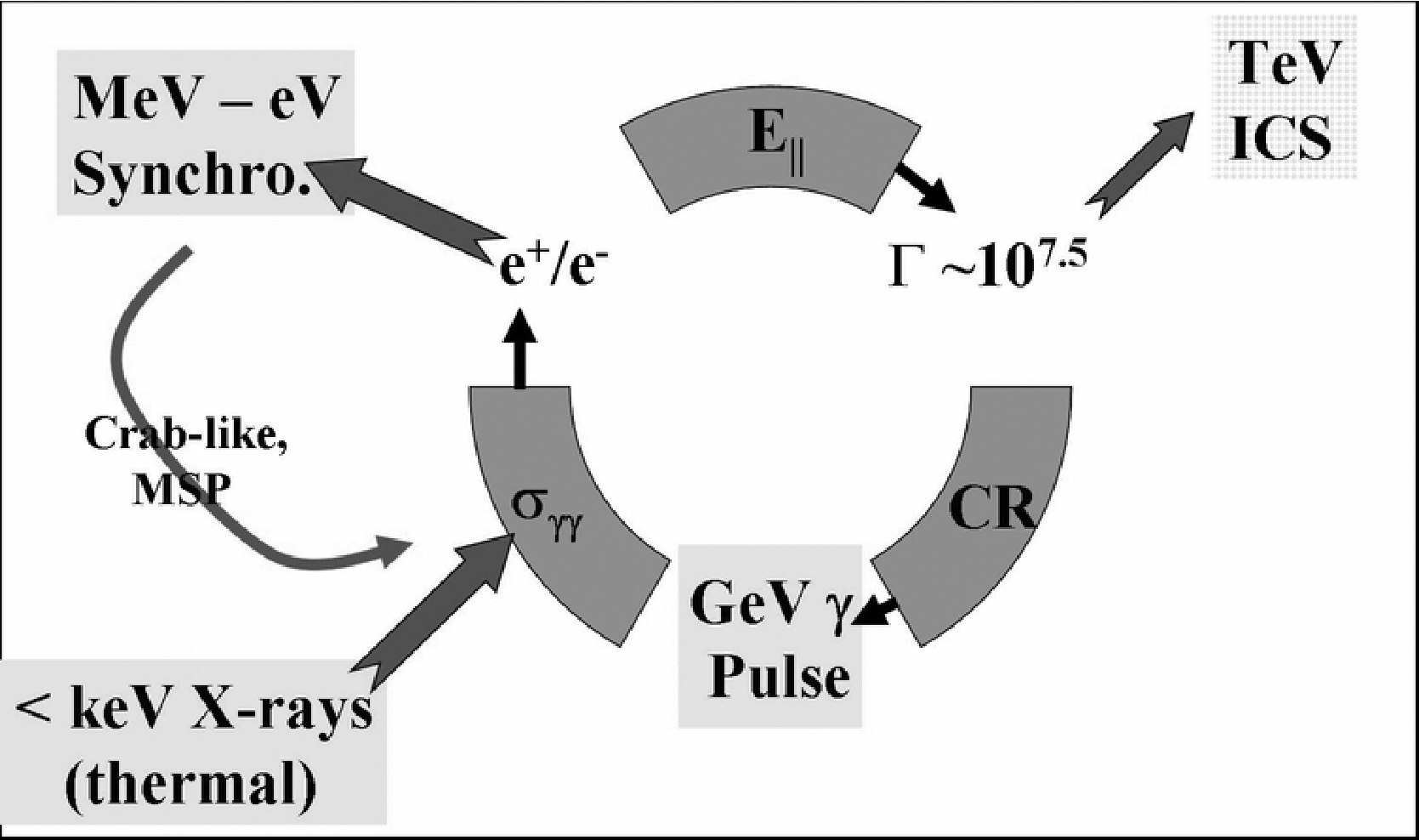}
\caption{Physical processes in outer magnetosphere gaps.}
\end{figure}

	In brief, acceleration of the primary electrons is limited by
curvature radiation reaction (CR). These curvature photons convert on
soft thermal (kT) or synchrotron (Sy) photons to make the pairs which
limit the growth of the acceleration gap. The thermal emission
produced by the general star surface and the pair-heated polar cap zone
may dominate the target soft photons for Vela-like pulsars, but for younger
pulsars and for MSP the synchrotron component should control the gaps.
This synchrotron emission comes from cooling of the initial pitch angle
of the $\gamma-\gamma$ pairs. Aside from the UV/soft X-ray band it dominates
the spectrum in the MeV-eV range. For Crab-like pulsars this emission may
be self-absorbed at sub-eV energies. There may also be IR-UV spectral features
related to the local cyclotron energy; in general we expect a peak in
the photon number flux in the optical/IR range. Interestingly, it seems
likely that this radiation represents an unusual low pitch angle
$\Psi \ll 1/\Gamma$ regime of synchrotron emission (Cruzius-W\"atzel,
Kunzl \& Lesch 2001). The presence of these soft photons guarantees some
Inverse Compton Scattering emission at multi-TeV energies, however the IR spectral
break of the synchrotron component and the substantial beaming of this
low energy flux greatly reduce the ICS optical depth and make computation
of the ICS flux difficult.

	Phase-resolved measurements of the spectral breaks provide the best test
of these models. For example, while polar cap models predict a very abrupt GeV
turnover from $\gamma-B$ absorption, in
the outer magnetosphere the GeV cutoff is the more
gradual rollover of the primary curvature spectrum for $\Gamma \sim
10^{7.5}$ $e^\pm$.  Existing EGRET data suggest mild (exponential) cut-offs
for Vela and Geminga, but a definitive answer requires GLAST.
The peak of the synchrotron spectrum is at
$E \sim 0.5 B_{12}^{5/2} (P/0.1s)^{-13/2} (r_i/0.1r_{LC})^{-2}$MeV for
$r_i \sim 0.1 r_{LC}$ near the null-charge surface, where 
$\Psi \sim r_i/r_{LC}$ induced by aberration. Measurement of this component
and its variation with pulse phase thus probes both the birth-site 
of the $\gamma-\gamma$ pairs and the field in the emission zone. Similarly
measurement of the optical/IR spectral breaks (and polarization behavior)
can probe the cooled population of $e^\pm$. However, except for
Crab-like pulsars, this will be a challenge, as the emission is
faint, typically $m_V > 25$ even for the nearest young objects.

\begin{figure}[b!]
\vspace*{14.7cm}
\includegraphics{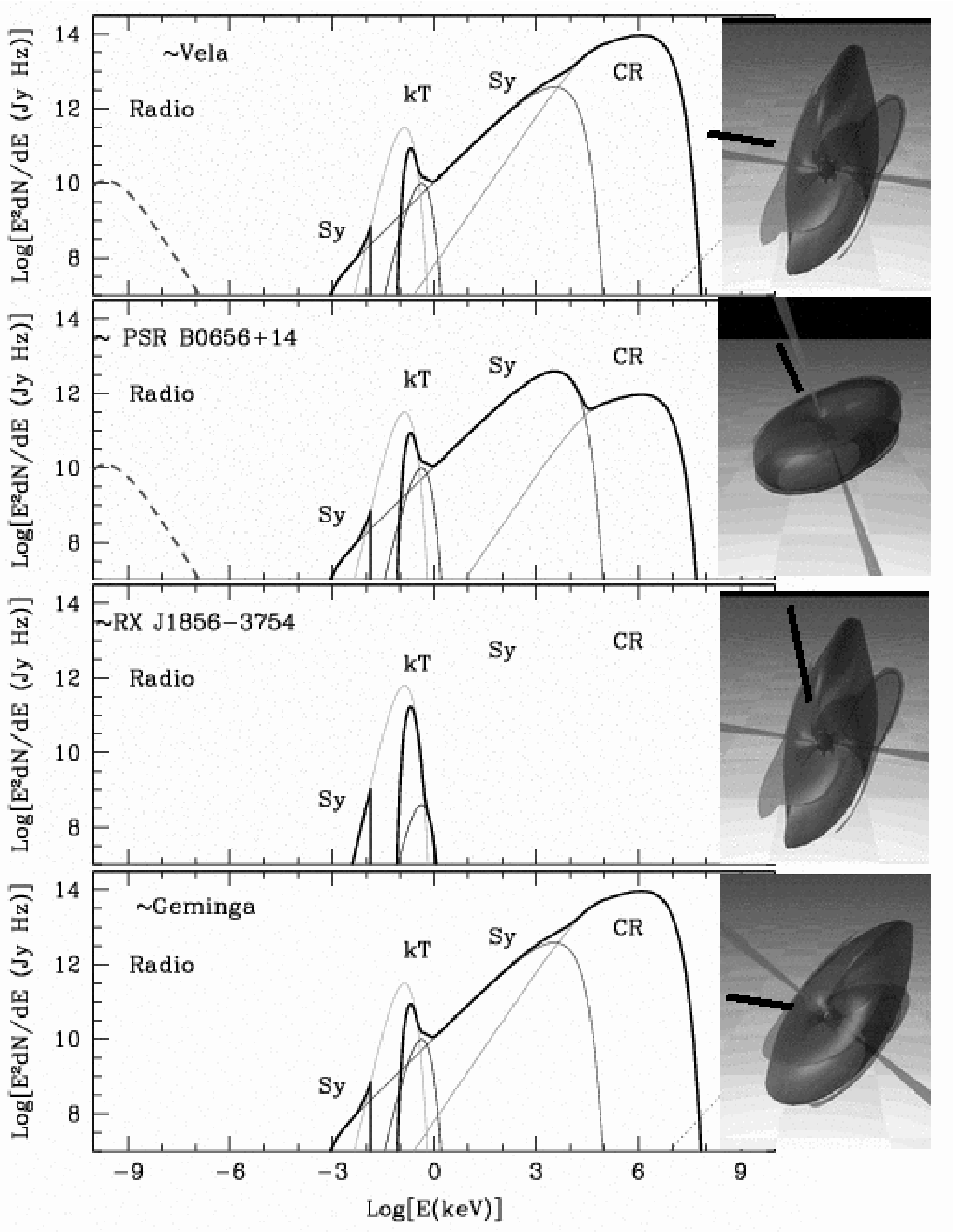}
\caption{Phase averaged pulsar SEDS. The line of sight is indicated in the
magnetosphere images to the right (black lines).
}
\end{figure}

\section{A `Unified' Beaming Model}

Of course,
the gap emissions described above are highly beamed, so the observed SED 
of a young, $\gamma$-ray active pulsar will depend  on the 
magnetic inclinations $\alpha$ and Earth viewing angle
$\zeta$. 
To illustrate how viewing geometry affects the source appearance, I present a
`unified model', in the spirit of the AGN unified model.
There are four main cases, illustrated in Figure 3. First, with
large $\alpha \approx \zeta \ga \pi/3$, the object is visible as a radio pulsar
and the active gaps dominate the high energy flux. A nearly isotropic 
thermal component (possibly with pulsations from a hot polar cap) is also
visible. The bulk of the {\it known} $\gamma$-ray pulsars fall in this class,
as the sample is highly biased to objects amenable to radio discovery.
The other class of easily discovered, radio bright young pulsars
has small $\alpha \approx \zeta$. For these objects the strong outer gap
CR component will miss Earth. GeV emission may be present, but it will
represent `off-pulse' emission with fluxes a small $\la 1$\% fraction of that
for the Vela-like objects. Evidence for such weak, more widely beamed flux
comes from the low level gap emission seen for Crab and Vela through all pulse 
phases. One picture for such `Off-Beam' emission in the polar cap
scenario has been described by Harding \& Zhang (2001). For small $\alpha 
\approx \zeta$ (e.g. PSR B0565+14) sources, the MeV synchrotron flux 
from the newly produced $e^\pm$ with relatively large $\Psi$ may still be
visible. Not also that for this geometry, we {\it expect} to see $\gamma$-ray
emission from near surface polar cap gaps, presumably with low flux.

	For $\alpha \ll \zeta$ neither the radio nor gap
beams will be visible, although the neutron star can still be discovered
via the nearly isotropic thermal emission. RX J1856-3754 provides
a likely example of this  geometry (Braje \& Romani 2002). Finally when
$\zeta$ is appreciably larger than $\alpha$, radio emission is not
visible, but the non-thermal gap emission will be strong. Geminga
is the archetype of this case. Since they lack radio pulses,
we expect that the handful of such objects known today is much smaller
than their true population.

Thus to understand individual SEDs, we need robust estimates of 
$\alpha$ and $\zeta$. The recent discoveries of equatorial tori in pulsar
wind nebula (PWN) provide such estimates for the Crab and Vela pulsars.
We have shown that even for fainter PWN useful constraints on $\zeta$
can be obtained (Romani \& Ng 2002). These measurements provide predictions of
which pulsars will show bright non-thermal high energy emission and can also
probe the relationship between neutron star linear and angular momenta
(Lai, these proceedings).

\section{Further Connections \& Conclusions}

	While geometry offers a uniform interpretation of the phase-averaged
SEDs of young neutron stars, as observation sensitivities increase, finer
details in the pulse profiles and phase-resolved spectra provide continued
challenges. For example Crab (Moffett \& Hankins 1996) and Vela
(Harding et al. 2002) multiwavelength profile compilations show at 
least four components each. Clearly multiple regions are active in the pulsar
magnetosphere. Of course, even if the bulk of the high energy emission
is produced in the outer magnetosphere, some pairs and faint
$\gamma$-rays associated
with the radio-generating polar cap gaps {\it must} be present. 
[Mutual gap poisoning may prevent both being visible from the same pole;
note that for outer gaps we automatically see the opposite pole in the radio.]
It seems plausible that with the $\sim 10^2-10^3 \times$ dynamic range 
available in the GLAST era, we may be able to identify $\gamma$
activity from both gaps.

	One further radio-high energy connection deserves comment. The Crab
giant radio pulses must, of course, be a high altitude phenomenon. The discovery
of similar pulses in millisecond pulsars, apparently aligned with strong
narrow X-ray components, suggests a more general connection between such pulses
and high energy emission (Romani \& Johnston 2001, Johnston, these proceedings).
With this zoo of behaviors, clearly the high-energy/radio connection
will keep observers and theorists busy for some time to come.


\begin{references}
\reference{} Braje, T.M. \& Romani, R.W. 2002, \apj in press.
\reference{} Cheng, K.S, Ho, C. \& Ruderman, M. 1986, \apj, 300, 522
\reference{} Cruzius-W\"atzel, A.R., Kunzl, T. \& Lesch, H. 2001, \apj, 546, 401
\reference{} Daugherty, J.K. \& Harding, A.K. 1996, \apj, 458, 278
\reference{} Harding, A.K., Musslimov, A.G. \& Zhang, B. 2002, \apj, 576, 366
\reference{} Harding, A.K. \& Zhang, B. 2001, \apjl, 548, L37
\reference{} Harding, A.K. et al. 2002, \apj, 576, 376
\reference{} Hirotani, K. \& Shibata, S. 1999, \mnras, 308, 54
\reference{} Moffett, D.A \& Hankins, T.H. 1996, \apj 468, 779
\reference{} Morini, M. 1983, \mnras, 202, 495
\reference{} Pavlov, G.G., Zavlin, V.E. \& Sanwal, D. 2002, astro-ph/0206024
\reference{} Romani, R.W. 1996, \apj, 470, 469
\reference{} Romani, R.W. \& Ng, C.-Y. 2002, \apj, submitted.
\reference{} Romani, R.W. \& Yadigaroglu, I.-A. 1995, \apj, 438, 314
\reference{} Romani, R.W. \& Johnston 2001, \apjl, 557, L93
\reference{} Thompson, D.J. 1998, in Neutron Stars and Pulsars, ed.
N. Shibazaki, N. Kawai, S. Shibata \& T. Kifune (Tokyo: Universal Academy) 273.
\reference{} Zhang, L. \& Cheng, K.S. 1997, \mnras, 487, 370

\end{references}
\end{document}